\documentclass[twocolumn,showpacs,preprintnumbers,amsmath,amssymb,pre]{revtex4}

\usepackage{graphicx}
\usepackage{dcolumn}
\usepackage{bm}
\usepackage{enumerate}

\begin{document}

\title{Random walk model of subdiffusion in a system with a thin membrane}

\author{Tadeusz Koszto{\l}owicz}
 \email{tadeusz.kosztolowicz@ujk.edu.pl}
 \affiliation{Institute of Physics, Jan Kochanowski University,\\
         ul. \'Swi\c{e}tokrzyska 15, 25-406 Kielce, Poland}

\date{\today}

\begin{abstract}
We consider in this paper subdiffusion in a system with a thin membrane. The subdiffusion parameters are the same in both parts of the system separated by the membrane. Using the random walk model with discrete time and space variables the probabilities (Green's functions) $P(x,t)$ describing a particle's random walk are found. The membrane, which can be asymmetrical, is characterized by the two probabilities of stopping a random walker by the membrane when it tries to pass through the membrane in both opposite directions. Green's functions are transformed to the system in which the variables are continuous, and then the membrane permeability coefficients are given by special formulae which involve the probabilities mentioned above. From the obtained Green's functions, we derive boundary conditions at the membrane. One of the conditions demands the continuity of a flux at the membrane but the other one is rather unexpected and contains the Riemann--Liouville fractional time derivative $P(x_N^-,t)=\lambda_1 P(x_N^+,t)+\lambda_2 \partial^{\alpha/2} P(x_N^+,t)/\partial t^{\alpha/2}$, where $\lambda_1$, $\lambda_2$ depending on membrane permeability coefficients ($\lambda_1=1$ for a symmetrical membrane), $\alpha$ is a subdiffusion parameter and $x_N$ is the position of the membrane. This boundary condition shows that the additional `memory effect', represented by the fractional derivative, is created by the membrane. This effect is also created by the membrane for a normal diffusion case in which $\alpha=1$.
\end{abstract}


\maketitle

\section{Introduction}\label{SecI}

Subdiffusion is usually defined as a random walk process in which $\left\langle (\Delta x)^2\right\rangle=2D_\alpha t^\alpha/\Gamma(1+\alpha)$, where $\left\langle (\Delta x)^2\right\rangle$ is a mean square displacement of a random walker, $\alpha$ is a subdiffusion parameter ($0<\alpha<1$), $D_\alpha$ is a subdiffusion coefficient and $\Gamma$ denotes the Gamma function \cite{mk}. The process of subdiffusion can occur in media in which the particles' movement is strongly hindered due to the internal structure of the medium, as, for example, in gels \cite{kdm}.
Subdiffusion is usually described by the following subdiffusion equation with the Riemann--Liouville fractional time derivative (here $0<\alpha<1$) \cite{mk}
\begin{equation}\label{eq44}
\frac{\partial C(x,t)}{\partial t}=D_\alpha\frac{\partial^{1-\alpha}}{\partial t^{1-\alpha}}\frac{\partial^2 C(x,t)}{\partial x^2}\;.
\end{equation}
The Riemann--Liouville derivative is defined as being valid for $\delta>0$ (here $k$ is a natural number which fulfils $k-1\leq \delta <k$)
\begin{equation}\label{eq45}
\frac{d^\delta f(t)}{dt^\delta}=\frac{1}{\Gamma(k-\delta)}\frac{d^k}{dt^k}\int_0^t{(t-t')^{k-\delta-1}f(t')dt'}\;.
\end{equation}
For $\alpha=1$ one obtains a normal diffusion equation. When one considers subdiffusion in a system with a thin membrane, which is treated as a partially permeable wall, a problem arises of how to set boundary conditions at the membrane \cite{gillespie,ibe,gardiner,korabel,kpre}.

The choice of boundary conditions at the membrane is a fundamental problem in the modelling of normal diffusion or subdiffusion in biological systems and in engineering science, when the filtration process is considered \cite{hobbie,luckey,hsieh}. In many papers, various boundary conditions at the membrane, related to specific processes of a particle's translocation through the membrane and in the vicinity of the membrane, have been used (see for example \cite{various}). In this paper we consider subdiffusion in a system in which a thin membrane seperates two homogeneous media having the same subdiffusion parameters. In such a system the most often used boundary condition at the membrane, whose complement demands that the flux is continuous at the membrane, reads $J(x_N^\pm,t)=b_1 C(x_N^-,t)-b_2 C(x_N^+,t)$ ($b_1=b_2$ for a symmetrical membrane) or $C(x_N^-,t)=kC(x_N^+,t)$ (see for example \cite{kpre,kjms} and the references cited therein), where $x_N$ is the membrane location, $J$ denotes a flux, $b_1$, $b_2$ and $k$ are parameters which control a membrane's permeability.
In the following, the above boundary conditions will be called the `old' boundary conditions.

We mention that diffusion in a system with a thick membrane can be treated as diffusion in a system with two partially permeable or partially absorbing walls, in which each of the membrane surfaces can be treated as such a wall \cite{kdl,kjms,shavit}.
However, in \cite{kdl} it is shown that the experimentally obtained concentration profiles coincide very well with the theoretical profiles if the latter fulfil the following boundary condition $C(x_N^-,t)=k(t)C(x_N^+,t)$, where ratio $k(t)$ changes exponentially over time to a constant value. To find the concentration profile for this case a special ansatz has been used, namely the constant parameter $k$ has been replaced by $k(t)$ in the solutions of the subdiffusion equation Eq. (\ref{eq44}) obtained previously for the above boundary condition with a constant $k$. However, such a procedure makes the functions obtained only approximately fulfil the subdiffusion equation. This remark shows that the above mentioned boudary condition cannot be treated as universal at a thin membrane when subdiffusion is described by Eq. (\ref{eq44}).
This fact is a major incentive finding a new boundary condition at a thin membrane in a system in which the membrane separates two homogeneous subdiffusive media which are characterized by the same subdiffusion parameters.

The methodology used in the consideration presented in this paper is as follows.
We use a simple random walk model of a particle in a system containing a partially permeable thin membrane in order to derive the Green's function $P(x,t;x_0)$ for this system. The Green's function can be interpreted as the probability density finding a particle at point $x$ after time $t$ under the condition that at the initial moment $t=0$ the particle was at the point $x_0$. This function can be also defined as a solution to the subdiffusion equation with appropriate boundary conditions for the following initial condition
\begin{equation}\label{eq0}
P(x,0;x_0)=\delta(x-x_0)\;, 
\end{equation}
where $\delta$ denotes the Dirac--delta function. When the Green's function is known, one can derive the boundary conditions at the wall. A similar methodology in which boundary conditions were derived from the Green's function was applied by Chandrasekhar \cite{chandrasekhar} to find boundary conditions at a fully absorbing or fully reflecting wall for a normal diffusion process.

The model used in this paper is based on a particle's random walk model on a discrete lattice. A discrete model of random walk appears to be a useful tool in modelling subdiffusion or normal diffusion \cite{gillespie,ibe,gardiner,weiss,chandrasekhar,hughes,hk,ks,kl2014}. We assume that the particle performs its single jump at a discrete time to at least the neighbouring site only. It is not permitted for a particle to stay at the site when the time of the jump is achieved, unless the particle is stopped by the wall with some probability, then the particle remains in its position. The particle's random walk in the membrane system is then described by a set of difference equations which can be solved by means of the generating function method \cite{weiss,hughes,montroll65,montroll64,barber}. Using the generating function obtained for these equations we pass from discrete to continuous time and space variables by means of the procedure presented in this paper. The choice of such a methodology is due to the fact that a diffusion process in a membrane system is relatively easy to model as a random walk in a discrete system.

\section{Model}\label{SecII}

We start our consideration with the difference equations which describe the random walk of a particle on a lattice.
Supposing $P_n(m;m_0)$ denotes the probability of finding a particle which has just arrived at site $m$ at the $n$--th step, $m_0$ is the initial position of the particle. The random walk is described by the following difference equation
\begin{equation}\label{eq1}
P_{n+1}(m;m_0)=\sum_{m'} p_{m,m'}P_n(m';m_0)\;,
\end{equation}
where $p_{m,m'}$ is the probability that a particle jumps from site $m'$ directly to site $m$. For subdiffusion or normal diffusion, long jumps can occur with a relatively small probability, thus we will take an often applied assumption \cite{chandrasekhar,weiss,hughes} that a jump can only be performed to neighbouring sites; it is not permissible to stay at the same site at the next moment unless a reflection from the wall occurs.

\begin{figure}
\includegraphics[scale=0.32]{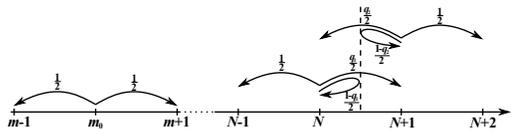}
\caption{System with a thin membrane located between $N$ and $N+1$ sites, a more detailed description is in the text.\label{Fig1}}
	\end{figure}

Let us suppose the membrane be located between the $N$ and $N+1$ sites. We assume that a particle which tries to pass through the membrane moving from the $N$ to $N+1$ site, can jump through the membrane with the probability $(1-q_1)/2$, but can be stopped by the membrane with the probability $q_1/2$, which means that the particle does not change its position after its `jump'. When the particle is located at site $N+1$, its jump to the $N$ can be performed with the probability $(1-q_2)/2$ and the probability that the particle can be stopped by the wall equals $q_2/2$ (see Fig.\ref{Fig1}).
The difference equation describing the random walk in a membrane system reads
\begin{eqnarray}
 \label{eq2}P_{n+1}(m;m_0)=\frac{1}{2}P_{n}(m-1;m_0)+\frac{1}{2}P_{n}(m+1;m_0),\nonumber\\ 
    m\neq N, N+1\;,\\
      \nonumber\\
  \label{eq3}P_{n+1}(N;m_0)=\frac{1}{2}P_{n}(N-1;m_0)+\frac{1-q_2}{2}P_{n}(N+1;m_0)\nonumber\\
  +\frac{q_1}{2}P_{n}(N;m_0)\;,\\
      \nonumber\\
  \label{eq4}P_{n+1}(N+1;m_0)=\frac{1-q_1}{2}P_{n}(N;m_0)+\frac{1}{2}P_{n}(N+2;m_0)\nonumber\\
  +\frac{q_2}{2}P_{n}(N+1;m_0)\;,
\end{eqnarray}
the initial condition is $P_0(m;m_0)=\delta_{m,m_0}$.
Since the probability of a particle's jump from the $N$ to $N+1$ sites equals $1/2$ in a system with a removed membrane, $q_1$ has an interpretation of a conditional probability of stopping a particle by the membrane under the condition that if the membrane were removed, the particle would take a jump from $N$ to $N+1$ site. Similar interpretation has the probability $q_2$.

In order to solve the Eqs. (\ref{eq2})--(\ref{eq4}) one uses a generating function method \cite{montroll64,barber}, with respect to a discrete time variable $n$, which is defined to be
\begin{equation}\label{eq5}
  S(m,z;m_0)=\sum_{n=0}^{\infty}z^nP_n(m,m_0)\;.
\end{equation}
From Eqs. (\ref{eq2})--(\ref{eq5}) we get the following difference equations
\begin{eqnarray}
  \label{eq6}\frac{1}{z}\left[S(m,z;m_0)-\delta_{m,m_0}\right]=\frac{1}{2}S(m-1,z;m_0)\\+\frac{1}{2}S(m+1,z;m_0)\;,\nonumber
  \qquad m\neq N, N+1\;,\\
     \nonumber\\
  \label{eq7}\frac{1}{z}\left[S(N,z;m_0)-\delta_{N,m_0}\right]=\frac{1}{2}S(N-1,z;m_0)\\+\frac{1-q_2}{2}S(N+1,z;m_0)\nonumber
  +\frac{q_1}{2}S(N,z;m_0)\;,\\
     \nonumber\\
  \label{eq8}\frac{1}{z}\left[S(N+1,z;m_0)-\delta_{N+1,m_0}\right]=\frac{1-q_1}{2}S(N,z;m_0)\\+\frac{1}{2}S(N-2,z;m_0)\nonumber
  +\frac{q_2}{2}S(N+1,z;m_0)\;.
\end{eqnarray} 
Solving Eqs. (\ref{eq6})--(\ref{eq8}) by means of the generating function method (the generating function $R$ for Eqs. (\ref{eq6})--(\ref{eq8}) is defined to be $R(u,z;m_0)=\sum_{m=-\infty}^\infty u^m S(m,z;m_0)$) we get
\begin{eqnarray}\label{eq9} 
& &S(m,z;m_0)=\frac{[\eta(z)]^{|m-m_0|}}{\sqrt{1-z^2}}\\
&+&\frac{\left([\eta(z)]^{|m-N|}-[\eta(z)]^{|m-N-1|}\right)}{\sqrt{1-z^2}}\nonumber\\
&\times &\frac{\eta(z)}{1-\eta(z)}\frac{\left(q_1[\eta(z)]^{|N-m_0|}-q_2[\eta(z)]^{|N+1-m_0|}\right)}{\left[1-(q_1+q_2-1)\eta(z)\right]}\nonumber\;,
\end{eqnarray}
where
\begin{equation}\label{eq10}
\eta(z)=\frac{1-\sqrt{1-z^2}}{z}\;.
\end{equation}
The particular forms of the function (\ref{eq9}), hereafter denoted as $S_{ij}$ where the indices $i,j$ denote the signs of $m-N,m_0-N$, respectively, are the following
\begin{eqnarray}
S_{--}(m,z;m_0)=\frac{[\eta(z)]^{|m-m_0|}}{\sqrt{1-z^2}}\label{eq11} \\
+\left[\frac{q_1-q_2\eta(z)}{1-(q_1+q_2-1)\eta(z)}\right]\frac{[\eta(z)]^{2N-m-m_0+1}}{\sqrt{1-z^2}}\;,\nonumber
\end{eqnarray}
\begin{equation}
S_{+-}(m,z;m_0)=\frac{[\eta(z)]^{m-m_0}(1+\eta(z))(1-q_1)}{\sqrt{1-z^2}\left[1-(q_1+q_2-1)\eta(z)\right]}\;,\label{eq12}
\end{equation}
\begin{equation}
S_{-\;+}(m,z;m_0)=\frac{[\eta(z)]^{m_0-m}(1+\eta(z))(1-q_2)}{\sqrt{1-z^2}\left[1-(q_1+q_2-1)\eta(z)\right]}\;,\label{eq13} 
\end{equation}
\begin{eqnarray}
S_{++}(m,z;m_0)=\frac{[\eta(z)]^{|m_0-m|}}{\sqrt{1-z^2}}\\
+\left[\frac{q_2-q_1\eta(z)}{1-(q_1+q_2-1)\eta(z)}\right]\frac{[\eta(z)]^{m+m_0-2N-1}}{\sqrt{1-z^2}}\;.\nonumber\label{eq14}
\end{eqnarray}

To pass from discrete to continuous time we use the following formula
\begin{equation}\label{eq15}
  P(m,t;m_0)=\sum_{n=0}^{\infty}P_n(m,m_0)\Phi_n(t)\;,
\end{equation}
where $\Phi_n(t)$ is the probability that the particle takes $n$ jumps in the time interval $(0,t)$.
In terms of the Laplace transform, $\mathcal{L}[f(t)]\equiv \hat{f}(s)=\int_0^\infty{{\rm e}^{-st}f(t)dt}$, the function $\Phi_{n}(t)$ reads \cite{mk}
\begin{equation}\label{eq16}
  \hat{\Phi}_n(s)=\frac{1-\hat{\omega}(s)}{s}\left[\hat{\omega}(s)\right]^n\;,
\end{equation}
where $\hat{\omega}(s)$ is the Laplace transform of a probability density $\omega(t)$ which is needed for the particle to take its next step.
Combining the Laplace transform of Eq. (\ref{eq15}) with Eqs. (\ref{eq5}) and (\ref{eq16}) we get
\begin{equation}\label{eq17} 
\hat{P}(m,s;m_0)=\frac{1-\hat{\omega}(s)}{s}S\left(m,\hat{\omega}(s);m_0\right)\;.
\end{equation}
We assume that $m_0\leq N$. From Eqs. (\ref{eq10}), (\ref{eq11}), (\ref{eq12}) and (\ref{eq17}) we obtain
\begin{eqnarray}\label{eq18}
\hat{P}_{--}(m,s;m_0)=\frac{1-\hat{\omega}(s)}{s\sqrt{1-[\hat{\omega}(s)]^2}}\Bigg[[\eta(\hat{\omega}(s))]^{|m-m_0|}\\
+\Lambda(s)[\eta(\hat{\omega}(s))]^{2N-m-m_0+1}\Bigg]\;,\nonumber
\end{eqnarray}
\begin{eqnarray}\label{eq19}
\hat{P}_{+-}(m,s;m_0)=\frac{1-\hat{\omega}(s)}{s\sqrt{1-[\hat{\omega}(s)]^2}}\\
\times M(s)[\eta(\hat{\omega}(s))]^{|m-m_0|}\;,\nonumber
\end{eqnarray}
where
\begin{equation}\label{eq20}
\Lambda(s)\equiv\frac{q_1-q_2\eta(\hat{\omega}(s))}{1-(q_1+q_2-1)\eta(\hat{\omega}(s))}\;,
\end{equation}
and
\begin{equation}\label{eq21}
M(s)\equiv \frac{(1-q_1)(1+\eta(\hat{\omega}(s)))}{1-(q_1+q_2-1)\eta(\hat{\omega}(s))}\;.
\end{equation}

Further considerations are performed assuming that $s$ is small, which corresponds to the case of large time due to Tauberian theorems.
For small $s$ there is
\begin{equation}\label{eq22}
\hat{\omega}(s)= 1-\tau_\alpha s^\alpha\;,
\end{equation}
where $\tau_\alpha$ is a parameter given in the units $1/s^\alpha$ which, together with $\alpha$, fully characterizes time distribution $\omega(t)$.
Eqs. (\ref{eq10}) and (\ref{eq22}) provide for small $s$
\begin{equation}\label{eq23}
\eta(\hat{\omega}(s))= 1-\sqrt{2\tau_\alpha s^\alpha}\;.
\end{equation}
From Eqs. (\ref{eq20}), (\ref{eq21}) and (\ref{eq23}) we obtain
\begin{eqnarray}
\Lambda(s)= \lambda_1+\lambda_2\sqrt{2\tau_\alpha s^\alpha}\;,\label{eq24}\\
    \nonumber\\
M(s)= 1-\lambda_1-\lambda_2\sqrt{2\tau_\alpha s^\alpha}\;,\label{eq25}
\end{eqnarray}
where
\begin{equation}\label{eq26}
\lambda_1=\frac{q_1-q_2}{2-q_1-q_2}\;,\;\lambda_2=\frac{(1-q_1)(q_1+q_2)}{(2-q_1-q_2)^2}\;.
\end{equation}

Supposing $\epsilon$ denotes the distance between discrete sites, and supposing 
\begin{equation}\label{eq27}
x=\epsilon m\;,\; x_0=\epsilon m_0\;,\;x_N=\epsilon N\;,
\end{equation} 
taking into consideration the following relation valid for small $\epsilon$
\begin{equation}\label{eq28}
\frac{P(m,t;m_0)}{\epsilon}\approx P(x,t;x_0)\;,
\end{equation}
and the following definition of the subdiffusion coefficient
\begin{equation}\label{eq29}
	D_\alpha=\frac{\epsilon^2}{2\tau_\alpha}\;,
\end{equation}
we pass from a discrete to a continuous space variable assuming $\epsilon\longrightarrow 0$.
However, a new problem arises within this limit. Namely, the mean number of steps $\left\langle n(t)\right\rangle$ over time interval $[0,t]$ is given by the following formula $\left\langle n(t)\right\rangle=\hat{\omega}(s)/[1-\hat{\omega}(s)]$, which together with Eqs. (\ref{eq22}) and (\ref{eq29}) provides $\left\langle n(t)\right\rangle\longrightarrow\infty$ when $\epsilon\longrightarrow 0$. Thus, for a very small $\epsilon$, $\left\langle n(t)\right\rangle$ takes anomalous large values. Then, the probability that a particle which tries to pass the partially permeable wall `infinite times' in every finite time interval passes through the wall, is equal to one. In order to avoid such a situation we assume that $q_1$ and $q_2$ are the functions of the parameter $\epsilon$ which fulfil $q_1(0)=q_2(0)=1$. For a small $\epsilon$ we suppose that
\begin{equation}\label{eq30}
q_1(\epsilon) = 1-\frac{\epsilon^\sigma}{\gamma_1}\;,\;q_2(\epsilon) = 1-\frac{\epsilon^\sigma}{\gamma_2}\;,
\end{equation}
where $\sigma$ is a positive number as yet to be determined, $\gamma_1$ and $\gamma_2$ are reflection membrane coefficients. From Eqs. (\ref{eq23})--(\ref{eq26}) and (\ref{eq30}), taking into account the relation (\ref{eq29}), we get
\begin{equation}\label{eq31}
\Lambda(s)=\frac{\gamma_1-\gamma_2}{\gamma_2+\gamma_1}+\frac{(2-w_\gamma\epsilon^\sigma)s^{\alpha/2}}{\gamma_1\sqrt{D_\alpha}w_\gamma^2}\epsilon^{1-\sigma}\;,
\end{equation}
\begin{equation}\label{eq32}
M(s)=\frac{2\gamma_2}{\gamma_2+\gamma_1}-\frac{(2-w_\gamma\epsilon^\sigma)s^{\alpha/2}}{\gamma_1\sqrt{D_\alpha}w_\gamma^2}\epsilon^{1-\sigma}\;,
\end{equation}
where $w_\gamma=1/\gamma_1+1/\gamma_2$.
The only value of $\sigma$ which ensures that the functions $\Lambda(s)$ and $M(s)$ are finite and depend on the wall's reflection parameters for any $\gamma_1$ and $\gamma_2$ (including the case of a symmetrical wall for which $\gamma_1=\gamma_2$) within the limit $\epsilon\longrightarrow 0$ is $\sigma=1$.
Thus, for small $\epsilon$ the reflection probabilities read
\begin{equation}\label{eq33}
q_1(\epsilon)= 1-\frac{\epsilon}{\gamma_1}\;,\;q_2(\epsilon)= 1-\frac{\epsilon}{\gamma_2}\;.
\end{equation}
Functions $q_1(\epsilon)$ and $q_2(\epsilon)$ are assumed to be decreased functions of $\epsilon$ and $q_1(\epsilon),q_2(\epsilon)\longrightarrow 0$ when $\epsilon\longrightarrow\infty$.
This seems to be a good choice
\begin{equation}\label{eq34}
q_1(\epsilon)={\rm e}^{-\frac{\epsilon}{\gamma_1}}\;,\;q_2(\epsilon)={\rm e}^{-\frac{\epsilon}{\gamma_2}}\;.
\end{equation}
These functions have the interpretation that a relative change of reflection probability is proportional to the change of distance of a particle from the wall, $dq_i(\epsilon)/q_i(\epsilon)=-d\epsilon/\gamma_i$, $i=1,2$.
From Eqs. (\ref{eq31}) and (\ref{eq32}) we obtain for $\sigma=1$ in the limit $\epsilon\longrightarrow 0$
\begin{equation}\label{eq35}
\Lambda(s)=\kappa_1+\kappa_2\frac{s^{\alpha/2}}{\sqrt{D_\alpha}}\;,
\end{equation}
\begin{equation}\label{eq36}
M(s)=1-\kappa_1-\kappa_2\frac{s^{\alpha/2}}{\sqrt{D_\alpha}}\;,
\end{equation}
where
\begin{equation}\label{eq37}
\kappa_1=\frac{\gamma_1-\gamma_2}{\gamma_1+\gamma_2}\;,\;\kappa_2=\frac{2/\gamma_1}{(1/\gamma_1+1/\gamma_2)^2}\;.
\end{equation}
The inverse formulae to (\ref{eq37}) read
\begin{equation}\label{eq38}
\gamma_1=\frac{2\kappa_2}{(1-\kappa_1)^2}\;,\;\gamma_2=\frac{2\kappa_2}{1-\kappa_1^2}\;.
\end{equation}
The parameters $\gamma_1$ and $\gamma_2$ have a relatively simple interpretation due to Eq. (\ref{eq34}). However, to shorten the notation, hereafter we will use the parameters $\kappa_1$ and $\kappa_2$ in the Green's functions and flux equations; the other equations and the description of the plots contain the parameters $\gamma_1$ and $\gamma_2$.
From Eqs. (\ref{eq18}), (\ref{eq19}), (\ref{eq22}), (\ref{eq27})--(\ref{eq29}) and (\ref{eq35})--(\ref{eq37}) we obtain in the limit $\epsilon\longrightarrow 0$
\begin{eqnarray}\label{eq39}
\hat{P}_{--}(x,s;x_0)=\frac{s^{-1+\alpha/2}}{2\sqrt{D_\alpha}}\Bigg[{\rm e}^{-\frac{|x-x_0|s^{\alpha/2}}{\sqrt{D_\alpha}}}+\\
+\Bigg(\kappa_1+\kappa_2\frac{s^{\alpha/2}}{\sqrt{D_\alpha}}\Bigg){\rm e}^{-\frac{(2x_N-x-x_0)s^{\alpha/2}}{\sqrt{D_\alpha}}}\Bigg]\nonumber\;,
\end{eqnarray}

\begin{eqnarray}\label{eq40}
\hat{P}_{+-}(x,s;x_0)=\frac{s^{-1+\alpha/2}}{2\sqrt{D_\alpha}}{\rm e}^{-\frac{(x-x_0)s^{\alpha/2}}{\sqrt{D_\alpha}}}\\
\times\Bigg(1-\kappa_1-\kappa_2\frac{s^{\alpha/2}}{\sqrt{D_\alpha}}\Bigg)
\;.\nonumber
\end{eqnarray}
Applying the following formula \cite{koszt}
\begin{eqnarray}\label{eq41}
\lefteqn{\mathcal{L}^{-1}\left[s^\nu {\rm e}^{-as^\beta}\right]\equiv f_{\nu,\beta}(t;a)=}\nonumber\\
&&=\frac{1}{t^{\nu+1}}\sum_{k=0}^\infty{\frac{1}{k!\Gamma(-k\beta-\nu)}\left(-\frac{a}{t^\beta}\right)^k}\;,
\end{eqnarray}
where $a,\beta>0$ (the function $f_{\nu,\beta}$ can be treated as a special case of the Fox function), the inverse Laplace transform of Eqs. (\ref{eq39}) and (\ref{eq40}) reads
\begin{eqnarray}\label{eq42}
P_{--}(x,t;x_0)=\frac{1}{2\sqrt{D_\alpha}}\Bigg[f_{\alpha/2-1,\alpha/2}\left(t;\frac{|x-x_0|}{\sqrt{D_\alpha}}\right)\\
+\kappa_1 f_{\alpha/2-1,\alpha/2}\left(t;\frac{2x_N-x-x_0}{\sqrt{D_\alpha}}\right)\Bigg]\nonumber\\
+ \frac{\kappa_2}{2D_\alpha}f_{\alpha-1,\alpha/2}\left(t;\frac{2x_N-x-x_0}{\sqrt{D_\alpha}}\right)\;,\nonumber
\end{eqnarray}
\begin{eqnarray}\label{eq43}
P_{+-}(x,t;x_0)=\frac{1-\kappa_1}{2\sqrt{D_\alpha}}\;f_{\alpha/2-1,\alpha/2}\left(t;\frac{x-x_0}{\sqrt{D_\alpha}}\right)\\
- \frac{\kappa_2}{2D_\alpha}\;f_{\alpha-1,\alpha/2}\left(t;\frac{x-x_0}{\sqrt{D_\alpha}}\right)\nonumber\;.
\end{eqnarray}

\section{Boundary conditions at the membrane}\label{SecIII}

The functions (\ref{eq42}) and (\ref{eq43}), which have been derived from the random walk model, fulfil the subdiffusion equation (\ref{eq44}). To check this, one can use the Laplace transform of Eq. (\ref{eq44}).
Since for $0<\delta<1$ and for a bounded function $f$ the Laplace transform of Eq. (\ref{eq45}) is
\begin{equation}\label{eq46}
\mathcal{L}\left[\frac{d^\delta}{d t^\delta}f(t)\right]=s^\delta\hat{f}(s)\;,
\end{equation}
the Laplace transform of Eq. (\ref{eq44}) reads
\begin{equation}\label{eq47}
s\hat{C}(x,s)-C(x,0)=s^{1-\alpha} D_\alpha\frac{\partial^2 \hat{C}(x,s)}{\partial x^2}\;.
\end{equation}
It is easy to check that the functions (\ref{eq39}) and (\ref{eq40}) fulfil Eq. (\ref{eq47}).

	\begin{figure}
\includegraphics[scale=0.32]{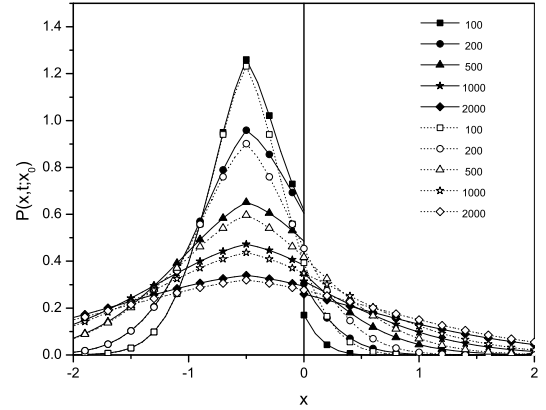}
\caption{Solid lines with filled symbols represent the functions (\ref{eq42}) and (\ref{eq43}), dotted lines with open symbols represent the functions (\ref{eq55}) and (\ref{eq56}) for $\gamma_1=\gamma_2=0.3$ and times given in the legend, the additional description is in the text.\label{Fig2}}
	\end{figure}

	\begin{figure}
\includegraphics[scale=0.32]{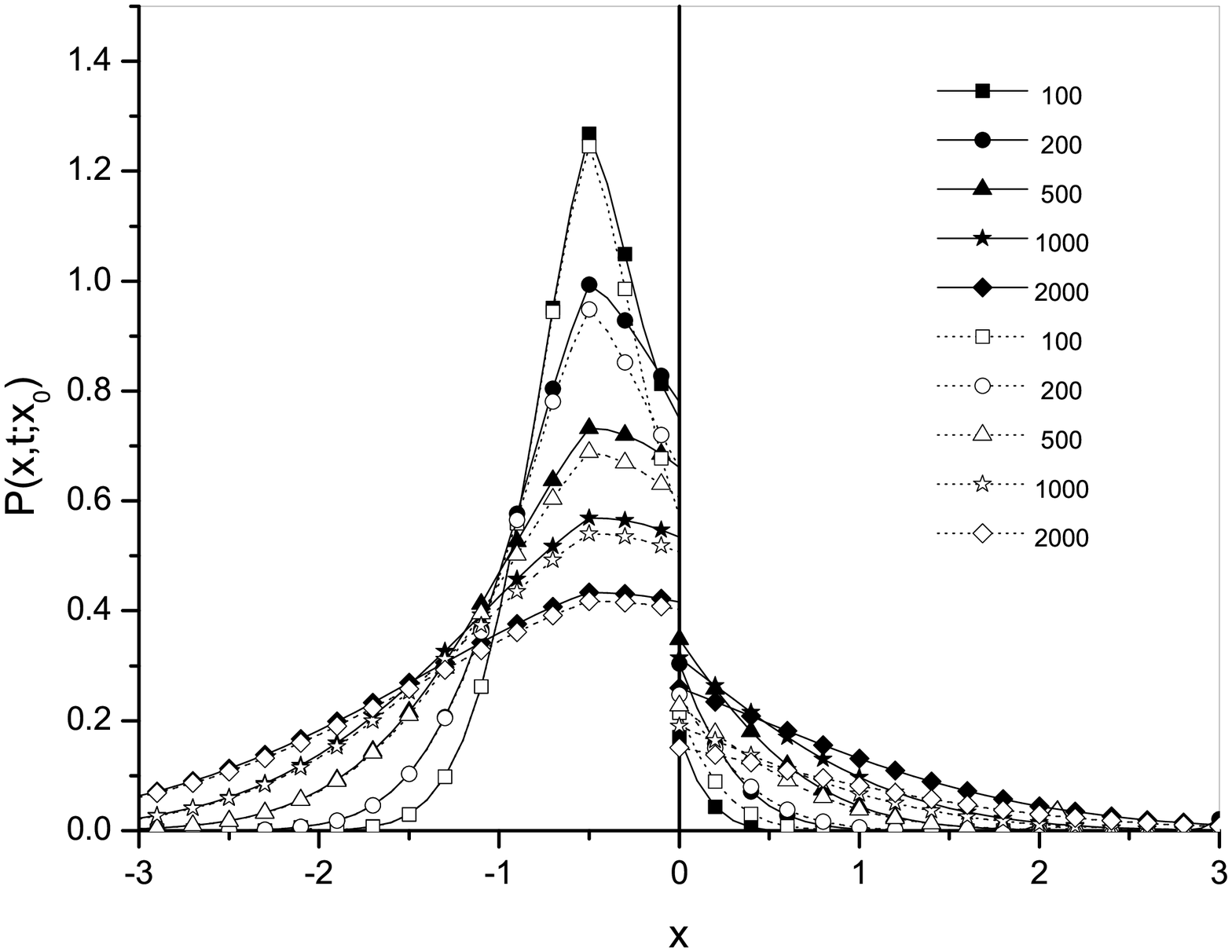}
\caption{Similar situation as in Fig. \ref{Fig2} but for $\gamma_1=0.8$, $\gamma_2=0.3$.\label{Fig3}}
	\end{figure}

	\begin{figure}
\includegraphics[scale=0.32]{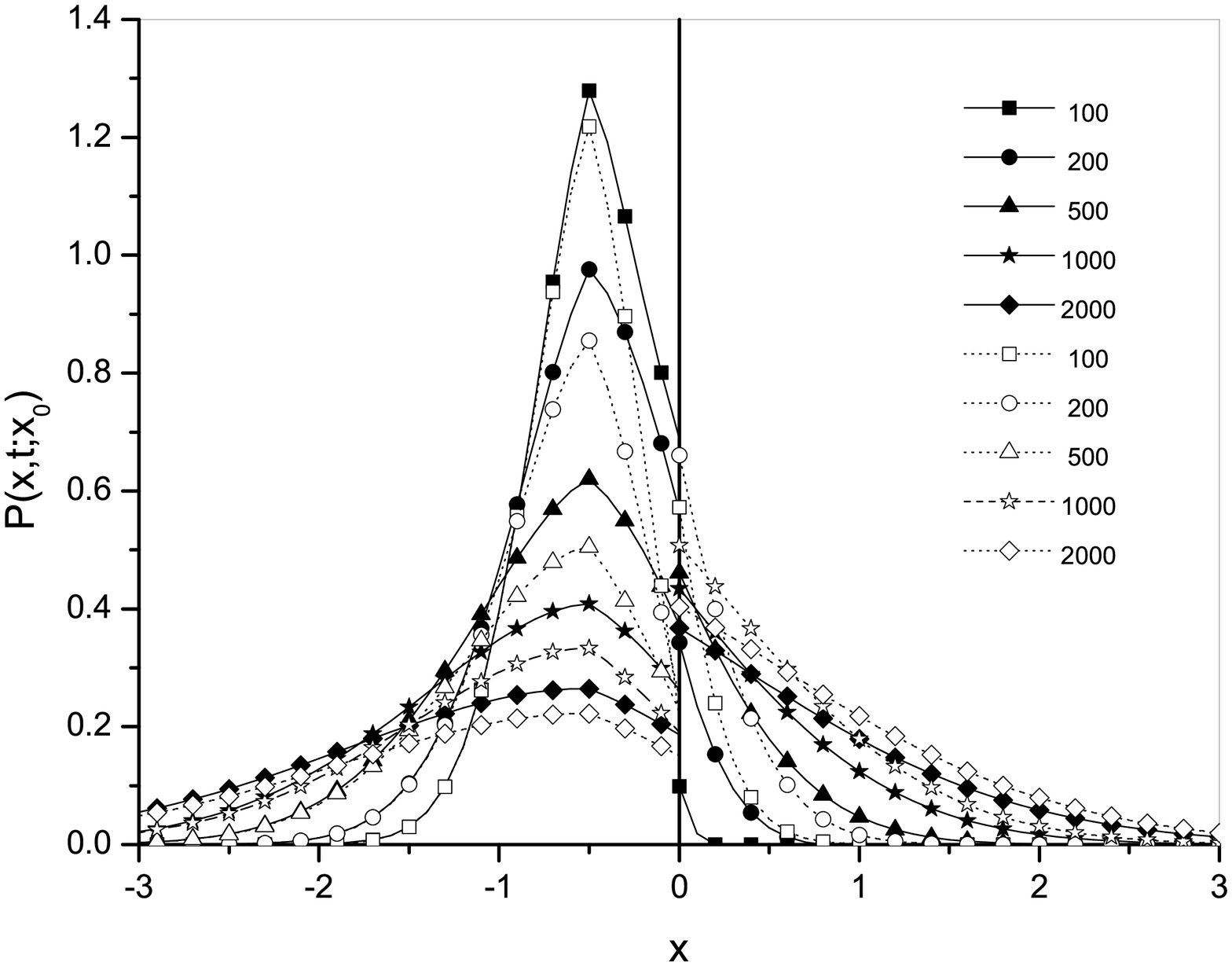}
\caption{Similar situation as in Fig. \ref{Fig2} but for $\gamma_1=0.3$, $\gamma_2=0.8$.\label{Fig4}}
	\end{figure}

	\begin{figure}
\includegraphics[scale=0.32]{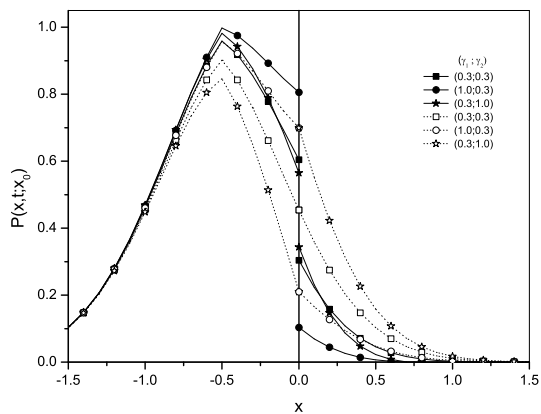}
\caption{Green's functions for various parameters $\gamma_1$ and $\gamma_2$ given in the legend, $t=200$.\label{Fig5}}
	\end{figure}
	
	\begin{figure}
\includegraphics[scale=0.32]{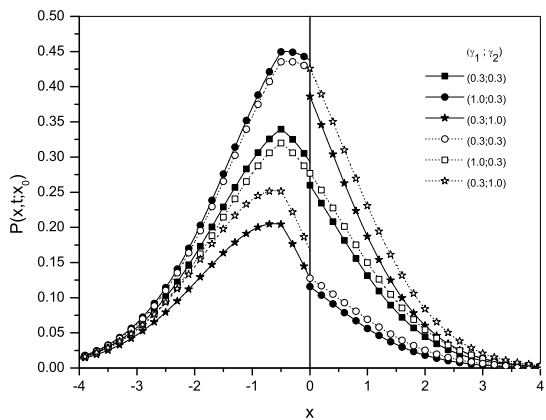}
\caption{Similar situation as in Fig. \ref{Fig5} but for $t=2000$.\label{Fig6}}
	\end{figure}

The subdiffusive flux generated by the solution to Eq. (\ref{eq44}), $C(x,t)$, is defined as
\begin{equation}\label{eq49}
J(x,t)=-D_\alpha\frac{\partial^{1-\alpha}}{\partial t^{1-\alpha}}\frac{\partial C(x,t)}{\partial x}\;.
\end{equation}
We note that combining this flux with the continuity equation $\partial C/\partial t=-\partial J/\partial x$ one gets the subdiffusion equation (\ref{eq44}).
Using (\ref{eq46}) we get the Laplace transform of the flux (\ref{eq49})
\begin{equation}\label{eq50}
\hat{J}(x,s)=-D_\alpha s^{1-\alpha}\frac{d \hat{C}(x,s)}{d x}\;.
\end{equation}
From Eqs. (\ref{eq39}), (\ref{eq40}) and (\ref{eq50}) we get
\begin{eqnarray}\label{flux1}
\hat{J}_{--}(x,s;x_0)=\frac{1}{2}\;{\rm sgn}(x-x_0)\;{\rm e}^{-\frac{|x-x_0|s^{\alpha/2}}{\sqrt{D_\alpha}}}\\
-\frac{1}{2}\left(\kappa_1+\kappa_2\frac{s^{\alpha/2}}{\sqrt{D_\alpha}}\right){\rm e}^{-\frac{(2x_N-x-x_0)s^{\alpha/2}}{\sqrt{D_\alpha}}}\nonumber\;,
\end{eqnarray}
\begin{eqnarray}\label{flux2}
\hat{J}_{+-}(x,s;x_0)=\frac{1}{2}\left(1-\kappa_1-\kappa_2\frac{s^{\alpha/2}}{\sqrt{D_\alpha}}\right){\rm e}^{-\frac{(x-x_0)s^{\alpha/2}}{\sqrt{D_\alpha}}}\;,
\end{eqnarray}
where the flux $J_{ij}$, $i,j=\pm$, is generated by the function $P_{ij}$, ${\rm sgn}$ denotes the signum function.
From Eqs. (\ref{eq39}), (\ref{eq40}), (\ref{flux1}) and (\ref{flux2}) we obtain the following Laplace transform of boundary conditions at the thin membrane
\begin{equation}\label{eq48}
\hat{P}_{--}(x_N,s;x_0)=\Bigg(\frac{\gamma_1}{\gamma_2}
+\gamma_1\frac{s^{\alpha/2}}{\sqrt{D_\alpha}}\Bigg)\hat{P}_{+-}(x_N,s;x_0)\;,
\end{equation}
\begin{equation}\label{eq51}
\hat{J}_{--}(x_N,s;x_0)=\hat{J}_{+-}(x_N,s;x_0)\;.
\end{equation}
The inverse Laplace transform of Eqs. (\ref{eq48}) and (\ref{eq51}) provides the following boundary conditions at a thin membrane
\begin{eqnarray}\label{eq52}
P_{--}(x_N,t;x_0)=\frac{\gamma_1}{\gamma_2}\;P_{+-}(x_N,t;x_0)\\ 
+\frac{\gamma_1}{\sqrt{D_\alpha}}\frac{\partial^{\alpha/2} P_{+-}(x_N,t;x_0)}{\partial t^{\alpha/2}}\;,\nonumber
\end{eqnarray}
\begin{equation}\label{eq53}
J_{--}(x_N,t;x_0)=J_{+-}(x_N,t;x_0)\;.
\end{equation}

The membrane's presence means that the passing of particles through the membrane should be treated as a process with a long memory (despite the fact that subdiffusion is the `long--memory' process itself), which is due to the last term in the right-hand side of Eq. (\ref{eq52}); this term is due to the presence of the last terms (containing $\kappa_2$) in Eqs. (\ref{eq39}), (\ref{eq40}) and (\ref{eq42}), (\ref{eq43}). Since a small $s$ corresponds to the large time $t$, the above mentioned terms can be omitted for a very large time, so the boundary condition (\ref{eq52}) takes the form
\begin{equation}\label{eq54}
P_{--}(x_N,t;x_0)=\frac{\gamma_1}{\gamma_2}\;P_{+-}(x_N,t;x_0)\;.
\end{equation}
It is easy to check that the solutions of the subdiffusion equation (\ref{eq44}) with the initial condition (\ref{eq0}) and boundary conditions (\ref{eq53}) and (\ref{eq54}) are
\begin{eqnarray}\label{eq55}
P_{--}(x,t;x_0)=\frac{1}{2\sqrt{D_\alpha}}\Bigg[f_{\alpha/2-1,\alpha/2}\left(t;\frac{|x-x_0|}{\sqrt{D_\alpha}}\right)\\
+\kappa_1 f_{\alpha/2-1,\alpha/2}\left(t;\frac{2x_N-x-x_0}{\sqrt{D_\alpha}}\right)\Bigg]\nonumber\;,
\end{eqnarray}
\begin{equation}\label{eq56}
P_{+-}(x,t;x_0)=\frac{1-\kappa_1}{2\sqrt{D_\alpha}}\;f_{\alpha/2-1,\alpha/2}\left(t;\frac{x-x_0}{\sqrt{D_\alpha}}\right)\;.
\end{equation}

The plots of Green's functions are presented in Fig. \ref{Fig2}--\ref{Fig6}. We focus our attention mainly on observing the difference between the functions (\ref{eq42}), (\ref{eq43}), represented by solid lines with filled symbols and (\ref{eq55}), (\ref{eq56}), represented by dotted lines with open symbols, the vertical line located in $x=0$ represents the membrane. In all cases we take $\alpha=0.9$, $D_\alpha=0.001$, $x_0=-0.5$, $x_N=0$, plots Fig. \ref{Fig2}--\ref{Fig4} are performed for various times given in the legend and in three cases concerning the relation between $\gamma_1$ and $\gamma_2$ parameters. In general, the values of the parameters and variables are given in arbitrarily chosen units. However, if we assume that the time unit is second and the space variable unit is a millimeter, then the parameter $D_\alpha$ is close to the real value of the subdiffuson coefficients of sugars in a water agarose solution \cite{kdm} for which $\alpha=0.9$. For the normal diffusion of ethanol in a water solvent in a system with artificial nucleopore membrane the values of $\gamma_1$ and $\gamma_2$ are of the order $10^{-1}\div 1$ \cite{kwl}. We assume that similar values of the nucleopore membrane reflecting coefficients are also valid for subdiffusive systems (the problem of extracting the membrane reflecting parameters from experimental data will be considered elsewhere \cite{kwl}). Therefore, we treat the values of parameters which are taken to calculation as quite realistic. In Fig. \ref{Fig2}--\ref{Fig6} we observe the differences between the Green functions (\ref{eq42}), (\ref{eq43}) and (\ref{eq55}), (\ref{eq56}). These differences, which decrease over a large time, are mostly visible in the near membrane region. For a symmetrical membrane (for which $\gamma_1=\gamma_2$ and $\kappa_1=0$) the `simplified' boundary condition Eq. (\ref{eq54}) provides the Green's function just like with the system without a membrane. This fact testifies that a `new' boundary condition at the membrane Eq. (\ref{eq52}) should be taken into account. The last term in brackets on the right-hand side of Eq. (\ref{eq48}) (and consequently in Eq. (\ref{eq52})) can be omitted if $\gamma_1/\gamma_2\gg \gamma_1 s^{\alpha/2}/\sqrt{D_\alpha}$. Since the condition $as^\beta \ll 1$ provides $t\gg a^{1/\beta}$, $a,\beta>0$, the boundary condition (\ref{eq54}) and the Green functions (\ref{eq55}) and (\ref{eq56}) are valid for $t\gg (\gamma_2/\sqrt{D_\alpha})^{2/\alpha}$. As far as we know, the membrane permeability coefficients $\gamma_1$ and $\gamma_2$ have not been considered for a subdiffusive system yet, at least in relation with an experiment, thus the estimation of time, for which Eqs. (\ref{eq54})--(\ref{eq56}) are valid, is unsure at this moment.
However, taking into account the `realistic' values of the parameters briefly discussed above, we note that for a time of the order of $10^4$ seconds Eqs. (\ref{eq42}), (\ref{eq43}) and (\ref{eq52}), should be used instead of Eqs. (\ref{eq54})--(\ref{eq56}). This conclusion is probably not universal, but we expect that for a lot of real membrane systems it works.

Using the procedure presented in this paper we obtain the following Green's functions for the case $x_0>x_N$
\begin{eqnarray}\label{eq42a}
P_{++}(x,t;x_0)=\frac{1}{2\sqrt{D_\alpha}}\Bigg[f_{\alpha/2-1,\alpha/2}\left(t;\frac{|x-x_0|}{\sqrt{D_\alpha}}\right)\\
-\kappa_1 f_{\alpha/2-1,\alpha/2}\left(t;\frac{x+x_0-2x_N}{\sqrt{D_\alpha}}\right)\Bigg]\nonumber\\
+ \frac{1+\kappa_1}{1-\kappa_1}\frac{\kappa_2}{2D_\alpha}f_{\alpha-1,\alpha/2}\left(t;\frac{x+x_0-2x_N}{\sqrt{D_\alpha}}\right)\;,\nonumber
\end{eqnarray}
\begin{eqnarray}\label{eq43a}
P_{-+}(x,t;x_0)=\frac{1+\kappa_1}{2\sqrt{D_\alpha}}\;f_{\alpha/2-1,\alpha/2}\left(t;\frac{x_0-x}{\sqrt{D_\alpha}}\right)\\
- \frac{1+\kappa_1}{1-\kappa_1}\frac{\kappa_2}{2D_\alpha}\;f_{\alpha-1,\alpha/2}\left(t;\frac{x_0-x}{\sqrt{D_\alpha}}\right)\nonumber\;.
\end{eqnarray}
The functions (\ref{eq42a}) and (\ref{eq43a}) fulfil the following boundary conditions
\begin{eqnarray}\label{eq57}
P_{++}(x_N,t;x_0)=\frac{\gamma_2}{\gamma_1}\;P_{-+}(x_N,t;x_0)\\ 
+\frac{\gamma_2}{\sqrt{D_\alpha}}\frac{\partial^{\alpha/2} P_{-+}(x_N,t;x_0)}{\partial t^{\alpha/2}}\;,\nonumber
\end{eqnarray}
\begin{equation}\label{eq58}
J_{++}(x_N,t;x_0)=J_{-+}(x_N,t;x_0)\;.
\end{equation}

Thus, one of the boundary conditions at the membrane depends on the position of the initial point $x_0$ and can be written in a compact form as follows
\begin{eqnarray}\label{eq59}
P_{\pm i}(x_N,t;x_0)=\frac{\gamma_i}{\tilde{\gamma}_{i}}\;P_{\mp i}(x_N,t;x_0)\\ 
+\frac{\gamma_i}{\sqrt{D_\alpha}}\frac{\partial^{\alpha/2}}{\partial t^{\alpha/2}}P_{\mp i}(x_N,t;x_0)\;,\nonumber
\end{eqnarray}
where $i=+,-$, $\gamma_i$ is the membrane reflection coefficient which controls the probability of stopping a particle by the membrane when a particle tries to pass the membrane from a region of its initial location to the opposite region, $\tilde{\gamma}_i$ is a membrane reflection coeffinient when the particle moves in the opposite direction; there is $\gamma_- =\gamma_1$, $\tilde{\gamma}_- =\gamma_2$ and $\gamma_+ =\gamma_2$, $\tilde{\gamma}_+ =\gamma_1$. In the next section, we present the procedure of involving the above boundary conditions in solving the subdiffusion equation for a case when an initial condition is given by an arbitrarily chosen function which can take non--zero values in both parts of the system.

\section{The method of solving a subdiffusion equation for an arbitrarily chosen initial condition}\label{SecIV}

As an example we solve the subdiffusion equation for the following initial condition
	\begin{equation}\label{eq69}
C(x,t)=\left\{
\begin{array}{ll}
C_0, & x<x_N,\\
0, & x>x_N.
\end{array}
\right.
	\end{equation}
We add that the initial condition (\ref{eq69}) often occurs when an experimental study of subdiffusion or normal diffusion in a membrane system are conducted, see \cite{kdm,kdl,dkmw} and the references cited therein.

	\begin{figure}
\includegraphics[scale=0.32]{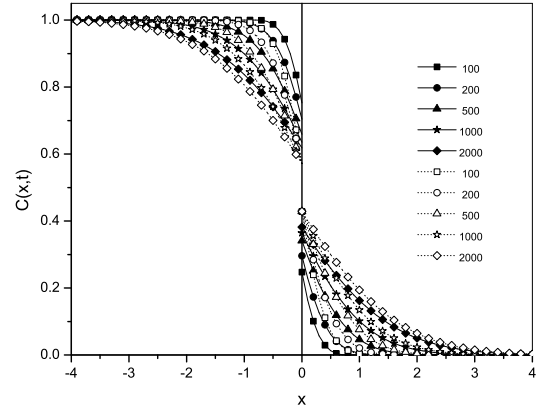}
\caption{Plots of the functions (\ref{eq70}), (\ref{eq71}) (solid lines with filled symbols) and (\ref{nr4}), (\ref{nr5}) (dashed lines with open symbols) for times given in the legend, $\gamma_1=0.4$, $\gamma_2=0.3$, $C_0=1$, the values of other parameters are the same as in previous figures.\label{Fig7}}
	\end{figure}

	\begin{figure}
\includegraphics[scale=0.32]{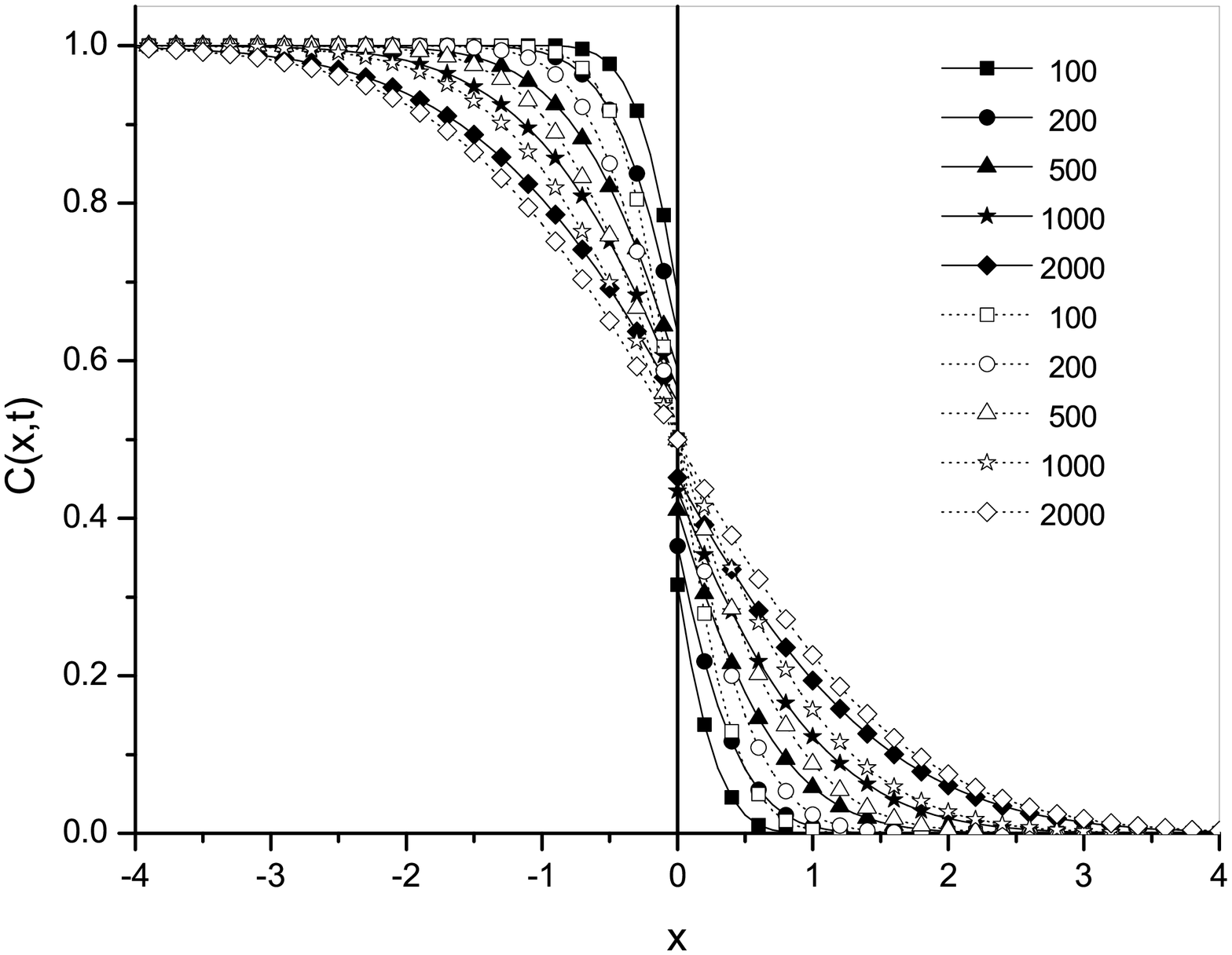}
\caption{Similar situation as in Fig. \ref{Fig7} but for $\gamma_1=\gamma_2=0.3$.\label{Fig8}}
	\end{figure}	

Let us denote 
\begin{equation}\label{eq60}
C(x,t)=\left\{
\begin{array}{ll}
C_-(x,t), & x<x_N,\\
C_+(x,t), & x>x_N.
\end{array}
\right.
	\end{equation}
Using the integral formula
\begin{equation}\label{eq63a}
C_{\pm}(x,t)=\int_{-\infty}^{x_N}{P_{\pm -}(x,t;x_0)C(x_0,0)dx_0}\;,
\end{equation}
from Eqs. (\ref{eq42}), (\ref{eq43}) and (\ref{eq69}) we obtain
	\begin{eqnarray}
C_-(x,t)&=&C_0\kappa_1+(1-\kappa_1)C_{F-}(x,t)\label{eq70}\\
&+&C_0\kappa_2 P_F(x,t;x_N)\;,\nonumber\\
 \nonumber\\
C_+(x,t)&=&(1-\kappa_1)C_{F+}(x,t)\label{eq71}\\
&-&C_0\kappa_2 P_F(x,t;x_N)\nonumber\;,
	\end{eqnarray}
where $P_F$ is the Green's function for the homogeneous system without a membrane
\begin{equation}\label{eq72}
P_F(x,t;x_0)=\frac{1}{2\sqrt{D_\alpha}}\;f_{\alpha/2-1,\alpha/2}\left(t;\frac{|x-x_0|}{\sqrt{D_\alpha}}\right)\;,
\end{equation}
$C_F$ denotes the solution to Eq. (\ref{eq44}) for the initial condition (\ref{eq69}) for the system without a membrane,
\begin{eqnarray}\label{eq73}
C_{F-}(x,t)=C_0-\frac{C_0}{2}f_{-1,\alpha/2}\left(t;\frac{x_N-x}{\sqrt{D_\alpha}}\right)\;,
\end{eqnarray}
for $x<x_N$ and
\begin{eqnarray}\label{eq74}
C_{F+}(x,t)=\frac{C_0}{2}f_{-1,\alpha/2}\left(t;\frac{x-x_N}{\sqrt{D_\alpha}}\right)\;,
\end{eqnarray}
for $x>x_N$.
The functions (\ref{eq70}) and (\ref{eq71}) fulfil the following boundary conditions at the membrane
\begin{eqnarray}\label{nr1}
C_{-}(x_N,t)&=&\frac{\gamma_1}{\gamma_2}C_{+}(x_N,t)\nonumber\\
&+&\frac{\gamma_1}{\sqrt{D_\alpha}}\frac{\partial^{\alpha/2}C_{+}(x_N,t)}{\partial t^{\alpha/2}}\;,
\end{eqnarray}
\begin{equation}\label{nr2}
J_{-}(x_N,t)=J_{+}(x_N,t)\;,
\end{equation}
where $J_\pm$ is the flux generated by $C_\pm$.
The boundary conditions Eqs. (\ref{nr1}) and (\ref{nr2}) can be obtained combining Eqs. (\ref{eq52}), (\ref{eq53}) and Eq. (\ref{eq63a}). As in the case of the boundary condition (\ref{eq52}), over a long time the boundary condition (\ref{nr1}) takes the following form
\begin{equation}\label{nr3}
C_{-}(x_N,t)=\frac{\gamma_1}{\gamma_2}C_{+}(x_N,t)\;.
\end{equation}
The solutions of the subdiffusion equation for the boundary conditions (\ref{nr2}) and (\ref{nr3}) read
	\begin{eqnarray}
C_-(x,t)&=&C_0\kappa_1+(1-\kappa_1)C_{F-}(x,t)\;,\label{nr4}\\
C_+(x,t)&=&(1-\kappa_1)C_{F+}(x,t)\;.\label{nr5}
	\end{eqnarray}
The plots of functions (\ref{eq70}), (\ref{eq71}) and (\ref{nr4}), (\ref{nr5}) are presented in Fig. \ref{Fig7} and Fig. \ref{Fig8}. It seems that the differences between the solutions obtained for the boundary condition (\ref{nr1}) and the ones obtained for (\ref{nr3}) are larger than the differences observed between the relevant Green's functions presented in Figs. \ref{Fig2}--\ref{Fig5}, especially over a long time.

The boundary condition (\ref{eq52}) is valid when $x_0<x_N$, similarly (\ref{eq57}) is valid when $x_0>x_N$. It is not possible to find a compact form of the boundary condition for an initial condition which is non-zero in both parts of the system separated by the membrane. In order to find the solution $C(x,t)$ of the subdiffusion equation for an arbitrarily chosen initial condition, we will find solutions for the above mentioned parts of the system separately.
The functions $C_-$ is a superposition of a particle's concentration $C_{-n}$ generated by the particles initially located in the interval $(-\infty,x_N)$ and a particle's concentration $C_{-p}$ generated by the particles initially located in the interval $(x_N,\infty)$; a similar superposition can be assumed for the function $C_+$. Thus, we have
\begin{eqnarray}
C_-(x,t)=C_{-n}(x,t)+C_{-p}(x,t)\;,\label{eq61}\\
C_+(x,t)=C_{+n}(x,t)+C_{+p}(x,t)\;,\label{eq62}
\end{eqnarray}
where
\begin{eqnarray}
C_{\pm n}(x,t)=\int_{-\infty}^{x_N}{P_{\pm-}(x,t;x_0)C_-(x_0,0)dx_0}\;,\label{eq63}\\
C_{\pm p}(x,t)=\int_{x_N}^{\infty}{P_{\pm+}(x,t;x_0)C_+(x_0,0)dx_0}\;.\label{eq64}
\end{eqnarray}
It is possible to find boundary conditions at the membrane for functions $C_{+n}$ and $C_{-n}$ alone, a similar remark concerns the functions $C_{+p}$ and $C_{-p}$. From Eqs. (\ref{eq52}), (\ref{eq53}) and (\ref{eq63}) we get
\begin{eqnarray}
C_{-n}(x_N,t)&=&\frac{\gamma_1}{\gamma_2}C_{+n}(x_N,t)\nonumber\\
&+&\frac{\gamma_1}{\sqrt{D_\alpha}}\frac{\partial^{\alpha/2}C_{+n}(x_N,t)}{\partial t^{\alpha/2}}\;,\label{eq65}\\
  \nonumber\\
J_{-n}(x_N,t)&=&J_{+n}(x_N,t)\label{eq66}\;,
\end{eqnarray}
and from Eqs. (\ref{eq57}), (\ref{eq58}) and (\ref{eq64}) we obtain
\begin{eqnarray}
C_{+p}(x_N,t)&=&\frac{\gamma_2}{\gamma_1}C_{-p}(x_N,t)\nonumber\\
&+&\frac{\gamma_2}{\sqrt{D_\alpha}}\frac{\partial^{\alpha/2}C_{-p}(x_N,t)}{\partial t^{\alpha/2}}\;,\label{eq67}\\
  \nonumber\\
J_{+p}(x_N,t)&=&J_{-p}(x_N,t)\label{eq68}\;.
\end{eqnarray}

Thus the procedure for solving the equation in a system with a thin membrane for an initial condition
\begin{equation}\label{nr6}
C(x,0)=\left\{
\begin{array}{ll}
C_-(x,0), & x<x_N,\\
C_+(x,0), & x>x_N,
\end{array}
\right.
	\end{equation}
can be presented at the following points.
\begin{enumerate}
	\item 
Find the following function
\begin{equation}\label{nr7}
C_n(x,t)=\left\{
\begin{array}{ll}
C_{-n}(x,t), & x<x_N,\\
C_{+n}(x,t), & x>x_N,
\end{array}
\right.
	\end{equation}
by solving the subdiffusion equation with the boundary conditions (\ref{eq65}), (\ref{eq66}) and the following initial condition	
\begin{equation}\label{nr8}
C_n(x,0)=\left\{
\begin{array}{ll}
C_-(x,0), & x<x_N,\\
0, & x>x_N.
\end{array}
\right.
	\end{equation}
	\item
Find the following function
\begin{equation}\label{nr9}
C_p(x,t)=\left\{
\begin{array}{ll}
C_{-p}(x,t), & x<x_N,\\
C_{+p}(x,t), & x>x_N,
\end{array}
\right.
	\end{equation}
by solving the subdiffusion equation with the boundary conditions (\ref{eq67}), (\ref{eq68}) and the following initial condition	
\begin{equation}\label{nr10}
C_p(x,0)=\left\{
\begin{array}{ll}
0, & x<x_N,\\
C_+(x,0), & x>x_N.
\end{array}
\right.
	\end{equation}
	
	\item

In order to obtain the solution of the subdiffusion equation, given in the form of Eq. (\ref{eq60}), use Eqs. (\ref{eq61}) and (\ref{eq62}).

\end{enumerate}

\section{Final remarks}\label{SecV}

In this paper we find the new boundary condition at a thin membrane which contains a fractional time derivative; this derivative is present also in the boundary condition at a thin membrane located in a system in which normal diffusion occurs. The `new' boundary condition Eq. (\ref{eq52}) can be obtained from the `old' boundary condition (\ref{eq54}) (which assumes the constant ratio of the solutions defined at both of the membrane surfaces), adding to the latter a term which contains a fractional time derivative, this additional term vanishes over a long time limit. We briefly describe the procedure of solving the subdiffusion equation when the `new' boundary condit ion is taken into consideration. We briefly discuss the time domain in which a `new' boundary condition should be used, as well as the differences between the solutions of subdiffusive equations obtained for both of the above mentioned boundary conditions. The main conlusions are the following. We also show that the random walk model appears to be a useful tool in modelling subdiffusion in a system with partially permeable walls. 

The boundary condition (\ref{eq52}) can be approximated by the `old' one (\ref{eq54}) over the limit of a very long time. For a smaller time, which -- as we argued earlier -- corresponds to a time of the order of $10^4 s$ for `typical' parameters, the additional term in the boundary condition should be taken into account. This term provides the occurence of a fractional derivative, which creates an additional `memory effect' on the membrane. This effect should evidently be taken into account for the case of symmetrical membrane; othervise, putting $\gamma_1=\gamma_2$ we get a function for the system withoud membrane. The differences bewteen the solutions obtained for `new' boundary condition (\ref{eq52}) and the ones obtained for the `old' boundary condition (\ref{eq54}), observed in the plots Figs. \ref{Fig2}--\ref{Fig8}, are similar for both a system with symmetrical membrane and a system with asymmetrical membrane. Thus, it seems to be reasonable to assume that the additional term in the boundary condition should also be taken into account for a system in which the membrane is asymmetrical.
   
The consideration presented in this paper is also valid for the normal diffusion case, substituting $\alpha=1$ in the above presented functions and equations. We obtain an unexpected result in that a fractional time derivative of the order $1/2$ is involved in the boundary condition at a thin membrane in a normal diffusion model. Thus, normal diffusion in a membrane system appears to be a process with a `long memory' which is created by the membrane.

The diffusion model used in this paper leads to some surprising results.
The crossing of the particle through the thin membrane turns out to be a process with a `long memory'. This process means the boundary condition at the membrane depends on the nature of the transport of molecules in the system, and is dependent on parameter $\alpha$.
The `long memory membrane effect' is not caused by the accumulation of a substance in a thin membrane. This effect is caused by molecules waiting an anomalously long time in the vicinity of the membrane to perform a jump through the membrane, it becomes negligible over a `very long time' limit.
This causes the Green's functions to have an interesting interpretation for some initial conditions. For example, for a system in which a homogeneous solution is separated by a membrane from a pure solvent at the initial moment, Eq. (\ref{eq69}), the solution of the subdiffusion equation is a combination of the solution of the equation for a system without a membrane (this part is controlled by parameter $\kappa_1$) and the Green's function obtained for a homogeneous system without a membrane which is interpreted as a source point of particles (positive or negative, see Eqs. (\ref{eq70}) and (\ref{eq71})) controlled by parameter $\kappa_2$.
We note that the membrane system described above is suitable for experimental study that can check the results presented in this work (as examples of similar experimental studies, see \cite{kdm,kdl,dkmw}).

\end{document}